\documentclass{article}
\usepackage{amsfonts}
\usepackage{CJK,amsmath,indentfirst}
\usepackage{graphics}
\usepackage{graphicx}
\usepackage{cite}
\begin{document}

\title{\bf Nanopteron solution of the Korteweg-de Vries equation}

\author{\footnotesize Jian-Yong Wang$^{1,2}$, Xiao-Yan Tang$^{1,3}$, S. Y. Lou$^{1,4,5}$, Xiao-Nan Gao$^{1}$, Man Jia$^{4}$\\
\footnotesize $^{1}$\it Department of Physics and Astronomy, Shanghai Jiao Tong University, Shanghai, 200240, China\\
\footnotesize $^{2}$\it Department of Mathematics and Physics, Quzhou University, Quzhou 324000, China\\
\footnotesize $^{3}$\it Institute of System Science, School of Information Science Technology, East China Normal University, Shanghai, 200241, China\\
\footnotesize $^{4}$\it Faculty of Science, Ningbo University, Ningbo, 315211, China\\
\footnotesize $^{5}$ \it Shanghai Key Laboratory of Trustworthy Computing, East China Normal University, Shanghai 200062, China}
\date{}
\maketitle
\parindent=0pt
{\bf Abstract:}  The nanopteron, which is a permanent but weakly
nonlocal soliton, has been an interesting topic in numerical study
for many decades. However, analytical solution of such a special
soliton is rarely considered. In this Letter, we study the explicit
nanopteron solution of the Korteweg-de Vries (KdV) equation.
Starting from the soliton-cnoidal wave solution of the KdV equation,
the nanopteron structure is shown to exist. It is found that for the
suitable choice of the wave parameters, the soliton core of the
soliton-cnoidal wave trends to be a classical soliton of the KdV
equation and the surrounded cnoidal periodic wave appears as small
amplitude sinusoidal variations on both sides of the main core. Some
interesting features of the wave propagation are revealed. In
addition to the elastic interaction, it is surprising that the phase
shift of the cnoidal periodic wave after the interaction with the
soliton core is always half of its wavelength, and this conclusion
is universal to soliton-cnoidal wave interactions.


\vskip.4in
\renewcommand{\thesection}{\arabic{section}}
\parindent=20pt

\section{Introduction}

The Korteweg-de Vries (KdV) equation
\begin{eqnarray}
u_t+Auu_x+Bu_{xxx}=0,\label{kdv} \label{kdv}
\end{eqnarray}
which was originally derived
to describe the propagation of gravity waves in shallow water \cite{kdv},
is now regarded as one of the most important systems in soliton theory.
It arises as a fundamental model in diverse branches of physics,
such as nonlinear optics, Bose-Einstein condensates and hydrodynamics \cite{application1}.
In particular, the KdV equation plays a significant role
in the study of small but finite amplitude
ion acoustic waves, magnetoacoustic waves, Alfv\'{e}n waves in plasma physics \cite{application2}.
It has been reported in an experimental observation that dynamical
properties of dust acoustic waves are found to agree quite well, particularly at low
amplitudes and low Mach numbers, with the classical soliton solution of the KdV equation \cite{experiment}.

Since the dramatic discovery of the particle-like  behavior of the localized waves by Zabusky and Kruskal in 1965 \cite{ZK},
there has been an unprecedented burst of research activities on solitons.
Several effective methods, such as the inverse scattering transformation method \cite{IST}, the Hirota bilinear formalism \cite{Hirota},
the Darboux transformation (DT) \cite{DT} , the B\"acklund transformation (BT) \cite{BT}, etc.,
have been developed to find the multiple soliton solutions of the KdV equation and other integrable systems.
Besides multiple soliton solutions, interactions between solitons and other types of nonlinear waves is another topic of great interest \cite{cnls,kdv1,nls,kdv2,bk}.
Recently, by combining the symmetry reduction method with the DT or BT related
nonlocal symmetries, researchers have established the interaction solutions between solitons and cnoidal periodic waves of
the KdV equation \cite{kdv1} as well as the nonlinear Schr\"odinger equation \cite{nls}.
Meanwhile, hinted by these results, two equivalent simple direct methods,
the truncated Painlev\'{e} and the generalized tanh function expansion approach,
are developed to find interaction solutions between solitons and other types of nonlinear waves,
such as cnoidal waves, Painlev\'e waves, Airy waves and Bessel waves \cite{kdv2,bk}.

In this Letter, we report a new analytical solution of a special
weakly nonlocal soliton of the KdV equation explicitly. Such a
particular solution is called nanopteron. The concept of  nanopteron
was originally introduced by Boyd when he was studying a weakly
nonlocal soliton in the $\phi^4$ model numerically. It is a
quasisoliton which almost satisfies the classical soliton, but fails
because of small amplitude oscillatory tails extending to infinity
in space \cite{boyd1}. During the past decades, the nanopteron
structure in both continuous and discrete systems has been studied
extensively
\cite{boyd1,boyd2,boyd3,hunter,topo1,nano1,nano2,nano3,nano4,nano5,nano6,nano7,nano8,nano9,nano10}.
For instance, Hunter and Scheurle have shown asymptotically that
capillary-gravity water waves can be consistently modeled by a
singularly perturbed KdV equation and solutions of this wave
equation are of nanopteron type when the Bond number is less than
one third \cite{hunter}. Actually, the investigation of interaction
between a topological soliton and a background small amplitude wave
has been an important topic in condensed matter physics for more
than three decades \cite{topo1}. In particular, it was shown that a
small amplitude oscillatory wave can propagate transparently through
a standing topological soliton with a phase shift. In addition, the
nanopteron structure has also been investigated in plasma physics
\cite{langmuir,keane}. Keane et al. studied the Alfv\'{e}n solitons
in a fermionic quantum plasma numerically \cite{keane}. Starting
from the governing equations for Hall magnetohydrodynamics including
quantum corrections, a coupled Zakharov-type system was derived and
numerically solved for both time independent and dependent cases.
The time-independent Alfv\'{e}n density soliton shares a similar
form of a nanopteron structure as an approximately Gaussian peak
surrounded by smaller sinusoidal variations. Then taking the
time-independent nanopteron solution as an initial condition, it was
numerically confirmed that the shape of the Gaussian peak retains
the same profile during its interaction with surrounded sinusoidal
variations. Obviously, some of the above results suggest that the
interaction between a soliton and a small amplitude background wave
is elastic, otherwise the moving waves will degenerate during their
propagations. Consequently, it is rather meaningful and significant
to obtain an analytical solution describing such types of waves.

\section{Soliton-cnoidal wave solution of the KdV equation}
Balancing the highest nonlinearity and dispersive terms in the KdV equation \eqref{kdv},
we assume its solution in the following generalized truncated $\tanh$ expansion
\begin{eqnarray}
u=u_0+u_1\tanh(w)+u_2\tanh(w)^2,\label{gtan}
\end{eqnarray}
where $u_0$, $u_1$, $u_2$ and $w$ are functions of $(x,t)$ to be determined later.

Substituting eq. \eqref{gtan} into eq. \eqref{kdv}
and vanishing all the coefficients of the different powers of $\tanh(w)$, we obtain the system of six overdetermined equations
that $u_0$, $u_1$, $u_2$ and $w$ need to satisfy.
It is fortunate to find that three of these over-determined equations are consistent.
From the coefficients of $\tanh(w)^5$, $\tanh(w)^4$ and $\tanh(w)^3$,
we find that $u_2$, $u_1$ and $u_0$ can be solved as
\begin{eqnarray}
u_2=-\frac {12Bw_x^2} {A},\label{ru2}
\end{eqnarray}
\begin{eqnarray}
u_1=\frac {12Bw_{xx}} {A},\label{ru1}
\end{eqnarray}
\begin{eqnarray}
u_0=\frac {B} {A}\left(\frac {3w_{xx}^2} {w_x^2}
-\frac {4w_{xxx}} {w_x}+8w_x^2\right)-\frac {w_t} {Aw_x}.\label{ru0}
\end{eqnarray}
Consequently, the solution \eqref{gtan} can be reformed in terms of $w$
\begin{eqnarray}
u&=&\frac {B} {A} \Big[12(w_{xx}\tanh(w)
+w_{x}^2\tanh^2(w))-\frac {w_t} {B w_x}\nonumber \\&&
+\frac {4w_{xxx}} {w_x}
-\frac {3w_{xx}^2} {w_{x}^2}-8w_x^2\Big].\label{ruw}
\end{eqnarray}
Then, from the coefficient of $\tanh(w)^2$,
we obtain the associated compatibility condition of $w$
\begin{eqnarray}
\frac {w_{t}} {w_{x}}+B \left(\frac {w_{xxx}} {w_{x}}
-\frac {3w^2_{xx}} {2w^2_{x}}-2w_{x}^2\right)+\lambda=0,\label{kdvw}
\end{eqnarray}
where $\lambda$ is a constant of integration.
Finally, one can verify that the remaining two over-determined equations
obtained from vanishing the coefficients of  $\tanh(w)^1$ and $\tanh(w)^0$ are
identically satisfied by using Eqs. \eqref{ru2}, \eqref{ru1}, \eqref{ru0} and \eqref{kdvw}.

In order to find the interaction solution between a soliton and a cnoidal wave,
we make the following ansatz for the solution of eq. \eqref{kdvw}
\begin{eqnarray}
&&w=\xi+c_1\mbox {arctanh}(c_2 sn(\eta,m)), \nonumber \\&&
\xi=\frac {x-V_1t} {W_1},\quad
\eta=\frac {x-V_2t} {W_2},\label{eqw}
\end{eqnarray}
where $sn$ is the usual Jacobi elliptic sine function and the parameter $m$ is known as its modulus.
$V_1$ and $V_2$ are velocities of the soliton and its surrounded cnoidal wave, respectively.
$W_1$ and $W_2$ are quantities related to the soliton width and the conidal wavelength, respectively.

Substituting the ansatz \eqref{eqw} back into eqs.
\eqref{ruw} and \eqref{kdvw} and setting zero the coefficients of
the different powers of Jacobi elliptic functions, we obtain a group
of overdetermined equations of the wave parameters $\{m, V_1, V_2,
W_1, W_2, c_1, c_2, \lambda\}$. When solving these overdetermined
equations, if we take the elliptic modulus $m$, velocities $V_{1}$
and $V_{2}$ as arbitrary, a nontrivial solution of the other five
wave parameters $\{W_1, W_2, c_1, c_2, \lambda\}$ can be determined
as
\begin{eqnarray}
W_1=\sqrt{\frac {8B(1-m^2)} {V_1-V_2}},\quad
W_2=\sqrt{\frac {2B(1-m^2)} {V_1-V_2}},\nonumber
\end{eqnarray}
\begin{eqnarray}
&&\lambda=\frac {(m^2-5)V_1+(3m^2+1)V_2} {4(m^2-1)},\nonumber \\&&
c_1=\frac {\delta} {2},\quad c_2=m,\quad \delta^2=1. \label{para}
\end{eqnarray}

By combining eqs. \eqref{ruw}, \eqref{eqw} and \eqref{para}, the explicit soliton-cnoidal wave solution of
the KdV equation can be obtained as
\begin{eqnarray}
u&=&\frac {3(V_1-V_2)} {2AG^2}\Big[\frac {(m^2-1+2G)^2} {(m^2-1)}\tanh(w)^2\nonumber \\&&
-2\delta m S(m^2-1+2G)\tanh(w)+m^2-1\Big]\nonumber \\&&
-\frac {(m^2+7)V_1-(3m^2+5)V_2} {2A(m^2-1)},\label{solu}
\end{eqnarray}
with
\begin{eqnarray}
&&G=1-m^2S^2+\delta m C D,\nonumber \\&&
S\equiv sn(\eta,m),\quad C\equiv cn(\eta,m), \quad D\equiv dn(\eta,m).\nonumber
\end{eqnarray}

As pointed out in our previous paper \cite{nls},
the soliton-cnoidal wave can be viewed as a dressed soliton,
namely, a soliton is dressed by a cnoidal periodic wave.
Consequently, the soliton-cnoidal wave can be divided into two parts.
By taking the limit $\tanh(w)=\pm1$ in eq. \eqref{solu},
we obtain the cnoidal periodic wave part of the dressed soliton
\begin{eqnarray}
C_{L}&=&\frac{3(V_1-V_2)(1+\delta m S)(m^2-1+2G)} {AG^2}\nonumber \\&&
+\frac{(5-m^2)V_1+(3m^2-7)V_2} {2A(m^2-1)},\nonumber \\&&
x-V_1t<0,\label{cl}
\end{eqnarray}
and
\begin{eqnarray}
C_{R}&=&\frac{3(V_1-V_2)(1-\delta m S)(m^2-1+2G)} {AG^2}\nonumber \\&&
+\frac{(5-m^2)V_1+(3m^2-7)V_2} {2A(m^2-1)}.\nonumber \\&&
x-V_1t>0.\label{cr}
\end{eqnarray}
Correspondingly, the soliton part of the wave is
\begin{eqnarray}
S_{L}&=&u-C_L,\quad x-V_1t<0,\label{sl}
\end{eqnarray}
and
\begin{eqnarray}
S_R&=&u-C_R,\quad x-V_1t>0.\label{sr}
\end{eqnarray}

To illustrate the dressed structure more clearly, let us
look at some figures. Fig. 1(a) exhibits the soliton-conidal wave
structure of $u$ determined by eq. \eqref{solu} at $t=0$. Fig. 1(b)
and Fig. 1(c) reveal the related structures of the cnoidal periodic
wave and the soliton core of $u$, respectively. Obviously, the
superposition of Fig. 1(b) and Fig. 1(c) is just Fig. 1(a). It is
observed from Fig. 1(b) that apart from the soliton center, the
solution rapidly tends to a cnoidal periodic wave. It is clear from
Fig. 1(c) that after removing the periodic wave background from $u$,
the left is just a soliton structure given by eqs. \eqref{sl} and
\eqref{sr}. Fig. 1(d) shows an elastic overtaking collision process
between a soliton and a cnoidal wave where both are right-going and
the soliton is traveling faster. It can be concluded from Fig. 1(d)
that despite the cnoidal periodic wave is a delocalized structure,
in the space-time evolution of the soliton-cnoidal wave, every peak
of the conidal periodic wave elastically interacts with the soliton
core except for a phase shift. To plot Fig. 1, the selection of the
nonlinearity coefficient $A$ and the dispersion coefficient $B$ are
given by
\begin{eqnarray}
&&A= \frac {48619\sqrt {1973}\sqrt {1980}}{315778650}\simeq0.304,\nonumber \\&&
B=\frac {388110739\sqrt{1973}\sqrt{1980}}{464096952000}\simeq1.653 \label{cab},
\end{eqnarray}
which is derived from the KdV equation describing the propagation of ion acoustic waves.

\input epsf
\begin{figure}[tbh]
\begin{center}
\includegraphics[width=6cm]{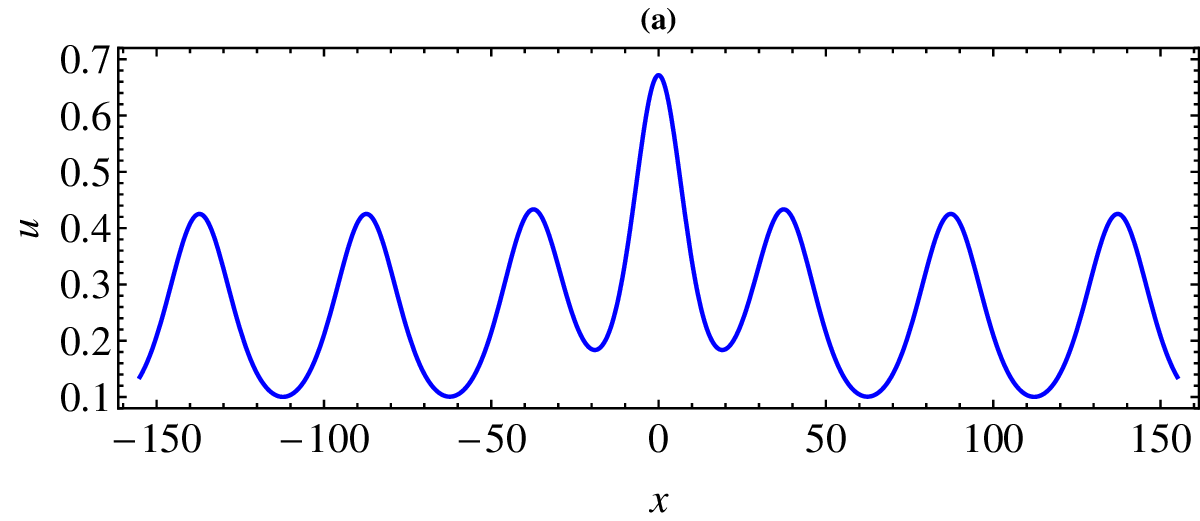}
\includegraphics[width=6cm]{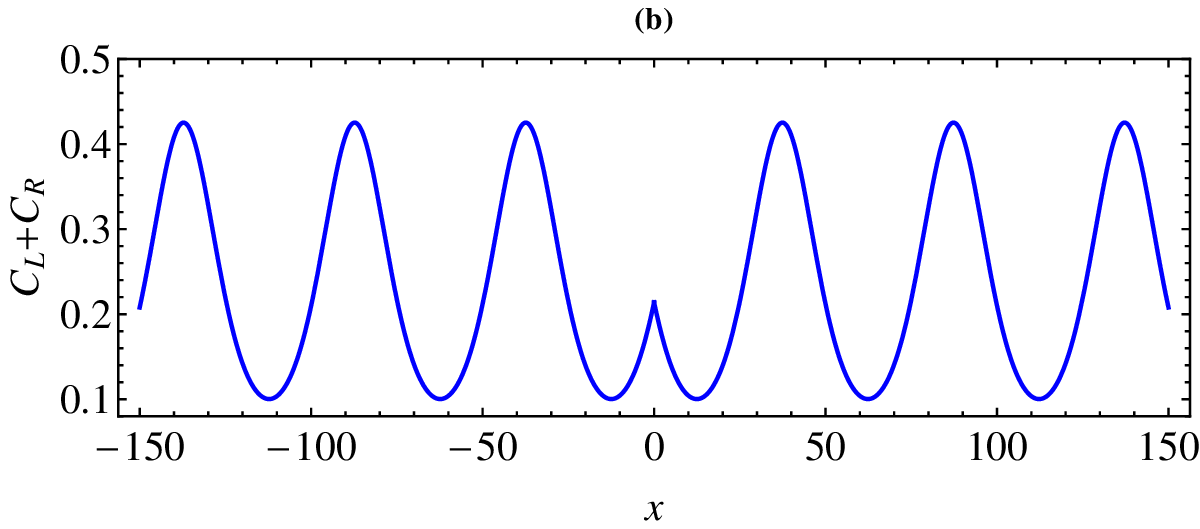}
\includegraphics[width=6cm]{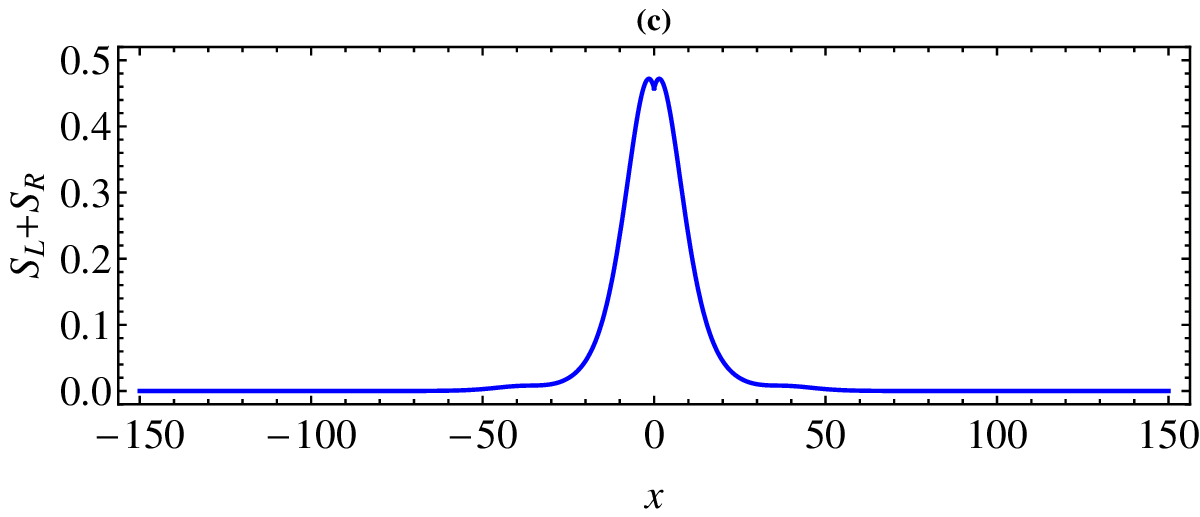}
\includegraphics[width=6cm]{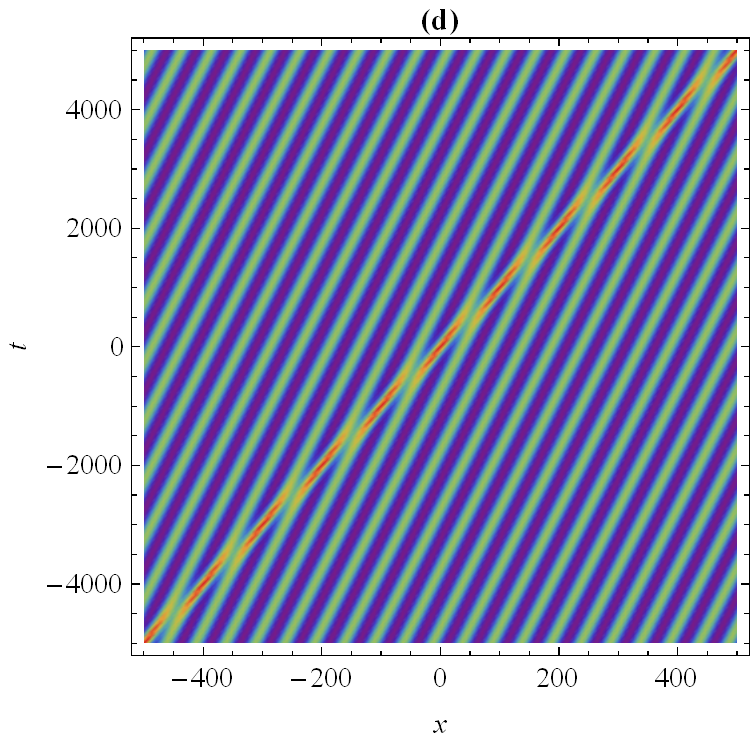}
\center{\footnotesize Fig. 1
(a) The soliton-cnoidal wave structure of the KdV equation given by eq. \eqref{solu}.
(b) The related cnoidal periodic wave structure given by eqs. \eqref{cl} and \eqref{cr}.
(c) The related soliton structure given by eqs. \eqref{sl} and \eqref{sr}.
(d) The density plot of $u$ for its space-time evolution.
The parameters are $m=0.3$, $V_1=0.1$, $V_2=0.05$, $\delta=1$, $A$ and $B$ are given by eq. \eqref{cab}.}
\end{center}
\end{figure}

The dressed structure enables us to compute the collision-induced phase shift of the cnoidal periodic wave.
For instance, now we consider an overtaking collision process as depicted in Fig. 1.
At time $t=0$, the conidal periodic wave peaks on the left side of the soliton core have interacted with
the soliton core, while not the cnoidal periodic peaks on the right side.
So there is a phase shift between them. Obviously, the conidal periodic wave peaks on the left and the right sides of the soliton core
can be expressed by $C_L$ and $C_R$, respectively.
By using eqs. \eqref{cl} and \eqref{cr}, it is surprising to find
that the collision-induced phase shift of the cnoidal periodic wave is given by
\begin{eqnarray}
\Delta_{cn}=2W_2K(m)=\frac {\lambda_c} 2,\label{shift}
\end{eqnarray}
where $\lambda_c$ is the wavelength of the cnoidal periodic wave, $K(m)$ is the first kind of complete elliptic integral.
This result can be easily verified. By substituting $\eta=\eta+2K(m)$ into \eqref{cl} and using the Jacobi elliptic identities
$sn(2K(m),m)=0$, $cn(2K(m),m)=-1$, and $dn(2K(m),m)=1$, we can directly demonstrate that  $C_{L}(\eta+2K(m))=C_{R}(\eta)$.
Similarly, by substituting $\eta=\eta+4K(m)$ into eqs. \eqref{cl}-\eqref{cr} and using the Jacobi elliptic identities
$sn(4K(m),m)=0$, $cn(4K(m),m)=1$, and $dn(4K(m),m)=1$, we can also demonstrate that  $C_{L}(\eta+4K(m))=C_{L}(\eta)$, and $C_{R}(\eta+4K(m))=C_{R}(\eta)$. So, $C_{L}$ and $C_{R}$ are functions of the period $4K(m)$.
Incidentally, the periods of the $sn$ and $cn$ functions are also $4K(m)$, while the period of the $dn$ function is $2K(m)$. It is noted that the phase shift of the interaction wave in Fig. 1(b) can be computed from eq. \eqref{shift} as 24.95, which coincides with the figure.

The phase shift formula \eqref{shift} tells that the phase shift of a cnoidal periodic wave after its interaction with a soliton is always half of its wavelength.
More significantly, this phase shift formula
is universal to all the soliton-cnoidal wave solutions obtained in Refs. \cite{nls,kdv1,kdv2}.
Unfortunately, it is still difficult to calculate the phase shift of the soliton because of the mixture of the $\tanh$ and Jacobi elliptic functions.
The existence of the Jacobi elliptic functions prevents us from calculating the difference of the
phases at two different time limits, approaching negative and positive infinities, respectively.
Therefore, an alternative method should be designed to overcome this difficulty.
\input epsf
\begin{figure}[tbh]
\begin{center}
\includegraphics[width=6cm]{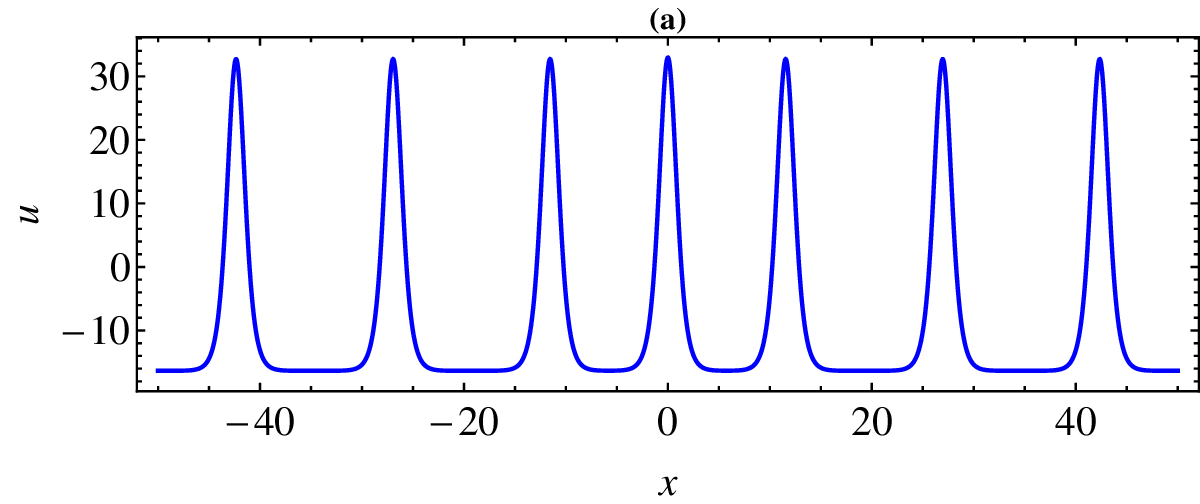}
\includegraphics[width=6cm]{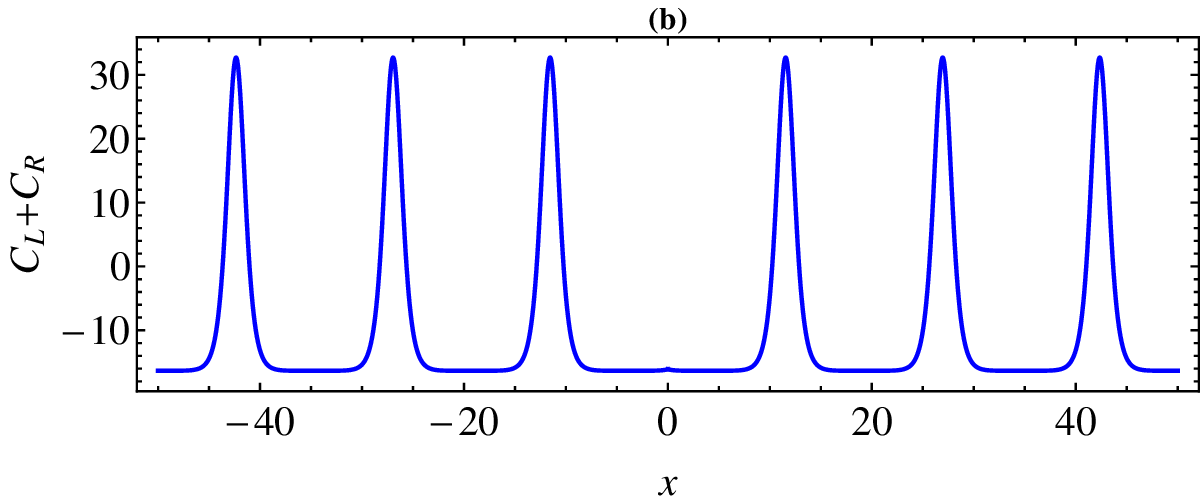}
\includegraphics[width=6cm]{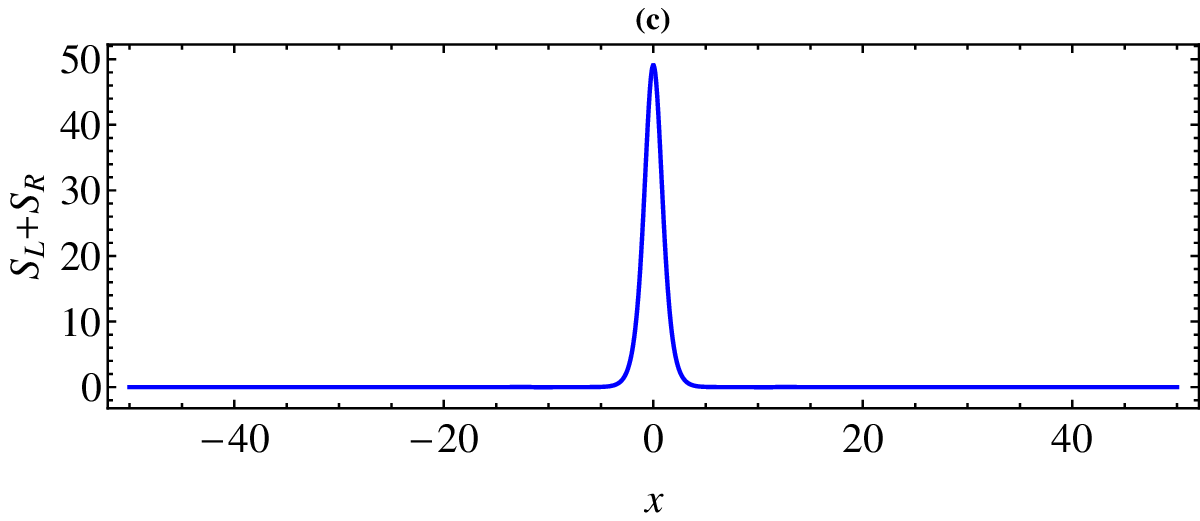}
\includegraphics[width=6cm]{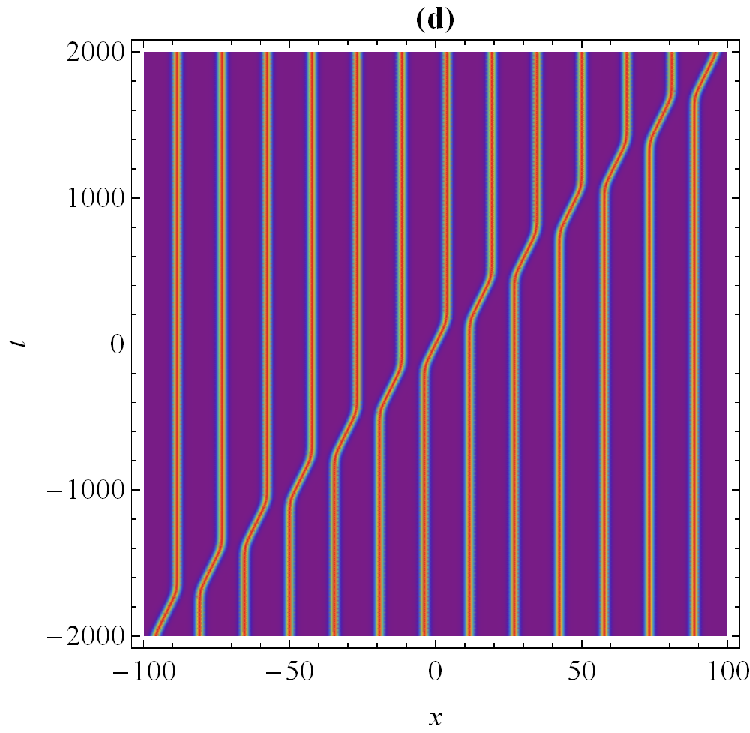}
\center{\footnotesize Fig. 2
(a) The soliton-cnoidal wave structure with $m=0.99$ at $t=0$.
(b) The related cnoidal periodic wave structure.
(c) The related soliton structure.
(d) The density plot of $u$ for its space-time evolution.
The parameters are $m=0.99$, $V_1=0.05$, $V_2=0$, $\delta=1$,
$A$ and $B$ are given by eq. \eqref{cab}.}
\end{center}
\end{figure}

Due to the fact that the parameter $m$ appears as not only the modulus of the Jacobi elliptic function
but also its coefficient since $c_2=m$, the amplitude of the cnoidal periodic wave trends to thrive with $m$
increasing. From Fig. 2(a), which is plotted to illustrate this phenomenon at $t=0$, it can be observed that
as soon as the parameter $m$ approaches $0.99$,
the amplitude of the cnoidal wave becomes comparable to the soliton core.
Fig. 2(b) shows that the solution \eqref{solu} exponentially approaches the cnoidal wave as $x\to\pm\infty$.
We also notice from Fig. 2(c) that after the periodic wave peaks $C_L$ and $C_R$ are taken away from the
exact solution $u$, only a tall and slim soliton structure is revived.
Fig. 2(d) reveals that the soliton core and every peak of the cnoidal periodic wave
can pass through each other transparently with a phase shift.

\section{Quasisoliton behavior as a nanopteron}
Before we proceed further,
let us first review the classical soliton solution of the KdV equation.
By using the usual $\tanh$ expansion method,
the single soliton solution of the KdV equation \eqref{kdv} can be obtained as
\begin{eqnarray}
u=\frac {1} {A W^2}[8B+VW^2-12B \tanh^2(\xi)],~ \xi=\frac {x-Vt} {W}.\label{sgs}
\end{eqnarray}
Imposing the boundary conditions
\begin{eqnarray}
u\to 0,\quad u_{\xi}\to 0,\quad u_{\xi\xi}\to 0,\quad \text{as}\quad \xi\to \pm\infty,\label{bound}
\end{eqnarray}
the width of the soliton can be determined and the above solution becomes
\begin{eqnarray}
u=\frac {3V} {A} {\rm sech}^2\left(\frac {x-Vt} {W}\right), \quad W=\sqrt{\frac {4B} {V}}.\label{cls}
\end{eqnarray}

Obviously, there is an intimate connection between the usual tanh function expansion method
and the generalized tanh function expansion method.
If we take $w$ as a straight line solution, namely, $w=(x-Vt)/W$,
the solution \eqref{ruw} reduces to the single soliton solution \eqref{sgs}
obtained by the usual tanh function expansion method.

Now, let us consider the asymptotic behavior of the soliton-cnoidal wave solution \eqref{solu}.
Under the ultra limit condtion $m=0$ ($G=1$), $V_1=V$ and $V_2=-V$, the wave parameters \eqref{para} degenerate to
\begin{eqnarray}
&&W_1=\sqrt{\frac {4B} {V}},\quad
W_2=\sqrt{\frac {B} {V}},\nonumber \\&&
\lambda=\frac {3V} {2},\quad
c_1=\frac {\delta} {2},\quad c_2=0,\label{paracla}
\end{eqnarray}
and the soliton-cnoidal wave solution \eqref{solu}
reduces to the classical soliton solution \eqref{cls}.
From eq. \eqref{paracla}, it is interesting to notice
that the substitution of the straight line solution $w=(x-Vt)/W$ into the compatibility condition \eqref{kdvw},
the width of the soliton can also be determined as $W=\sqrt{ 4B/V}$ with $\lambda=3V/2$.
However, the deeper physical reason why the compatibility condition of $w$ plays the similar role
as the boundary condition \eqref{bound} still need further consideration.
This interesting limit case hints us
to consider the asymptotic behavior of the soliton-cnoidal wave solution \eqref{solu}.
Under the asymptotic condition $V_1=V$, $V_2=-V$, and $m\to 0$,
we find that the soliton core profile goes to be the classical KdV soliton while
the surrounded conidal periodic wave becomes a small amplitude sinusoidal wave oscillating around zero.
This wave profile, in which the classical soliton is dressed by small amplitude oscillations,
is just the nanopteron structure proposed by Boyd.
In this sense, the soliton-cnoidal wave solution \eqref{solu} can also be named
as a nanopteron solution for the sake of its asymptotic behavior.

Under the limit $m\to0$, the function $K(m)$ trends to $\pi/2$.
Thus, the collision-induced phase shift of the small amplitude background wave can be approximately taken as
\begin{eqnarray}
\Delta_{sin}=2W_2K(m)\simeq W_2\pi.\label{shifts}
\end{eqnarray}

\input epsf
\begin{figure}[tbh]
\begin{center}
\includegraphics[width=10cm]{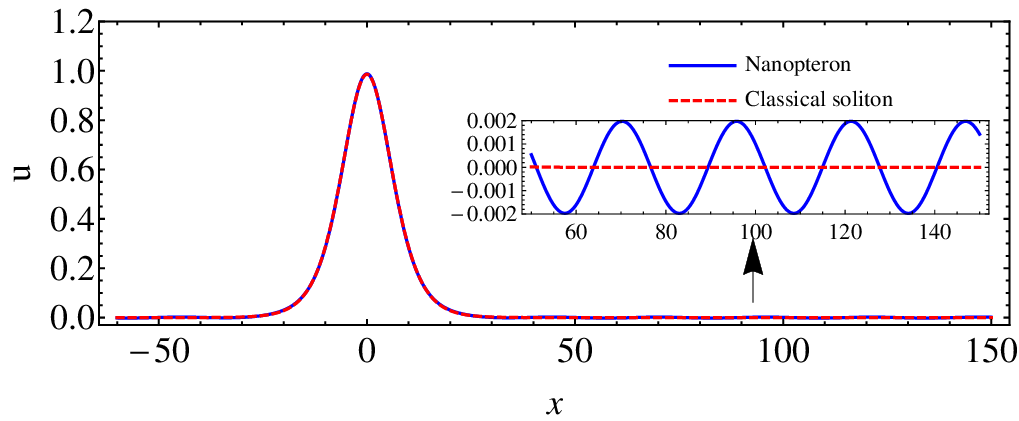}
\center{\footnotesize Fig. 3 A comparison of the classical soliton solution \eqref{cls} to the nanopteron solution \eqref{solu} at $t=0$.
The wave parameters are $V=0.1$, $V_1=0.1$, $V_2=-0.1$ $m=0.001$, $\delta=1$, $A$ and $B$ are given by eq. \eqref{cab}.}
\end{center}
\end{figure}

A comparison of the classical soliton to the nanopteron for $m=0.001$ at $t=0$ is given in Fig. 3, which demonstrates that the curves of two solutions coincide exactly with each other at a large space scale.
However, the inset on the right side of Fig. 3
shows that the oscillating tail is nonvanishing despite of a tiny amplitude.
When $m$ becomes a little larger, the nanopteron tail grows up conspicuously.
Fig. 4(a) presents a comparison of classical soliton to the nanopteron structures for $m=0.02$ and $\delta=\pm1$ at $t=0$.
It is observed that the soliton core of the nanopteron is higher and slightly narrower than the classical soliton when $\delta=-1$, while
shorter and slightly wider when $\delta=-1$.
Interestingly, the crests and troughs of the surrounded sinusoidal
waves of the nanopteron structures with $\delta=\pm1$ are corresponding to each other.
Fig. 4(b) and Fig. 4(c) reveal the dressed structure of the nanopteron solution with $\delta=1$.
Fig. 4(b) shows that apart from the soliton center,
the solution rapidly approaches a small amplitude sinusoidal wave oscillating around zero.
From eq. \eqref{shifts},
the phase shift of the small amplitude sinusoidal wave can be approximately calculated as $W_2\pi\simeq12.77$,
which is in accordance with Fig. 4(b).
Fig. 4(c) reveals that only a soliton $S_L+S_R$ is left after the sinusoidal wave are ruled out from the exact
solution $u$. Fig. 4(d) is a three-dimensional plot of the nanopteron solution with $\delta=1$.

\input epsf
\begin{figure}[tbh]
\begin{center}
\includegraphics[width=6cm]{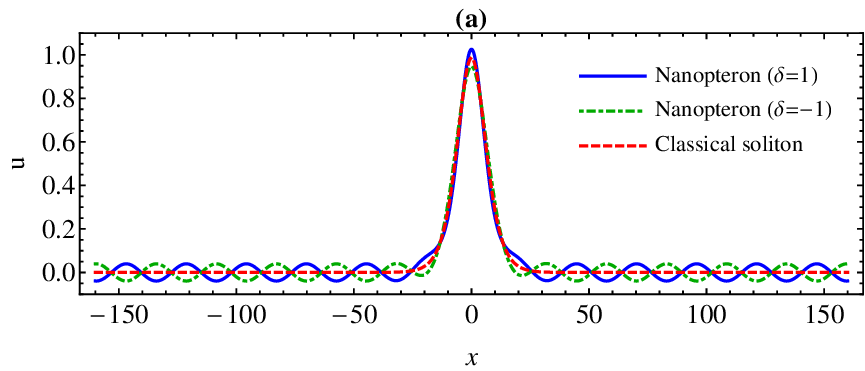}
\includegraphics[width=6cm]{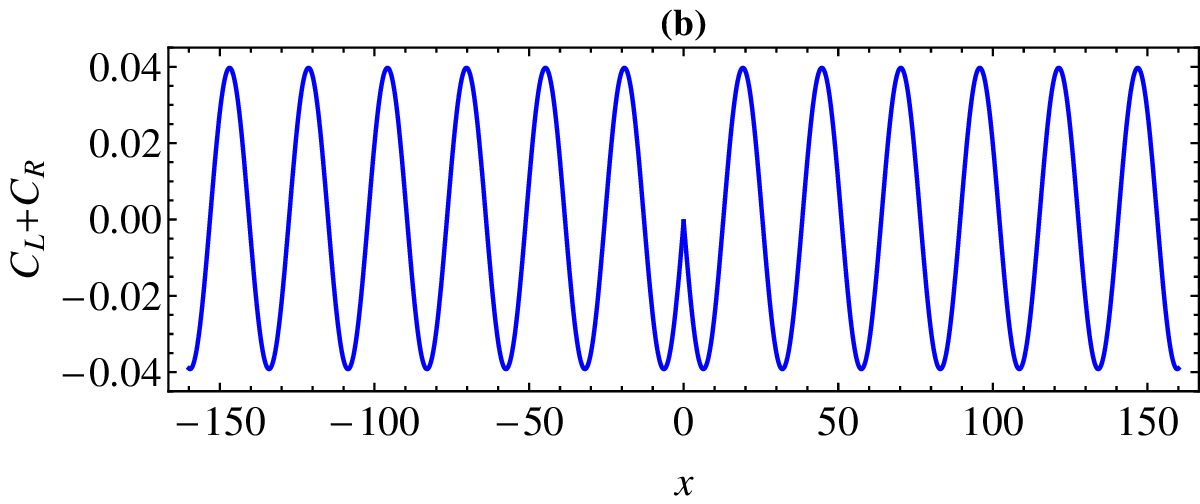}
\includegraphics[width=6cm]{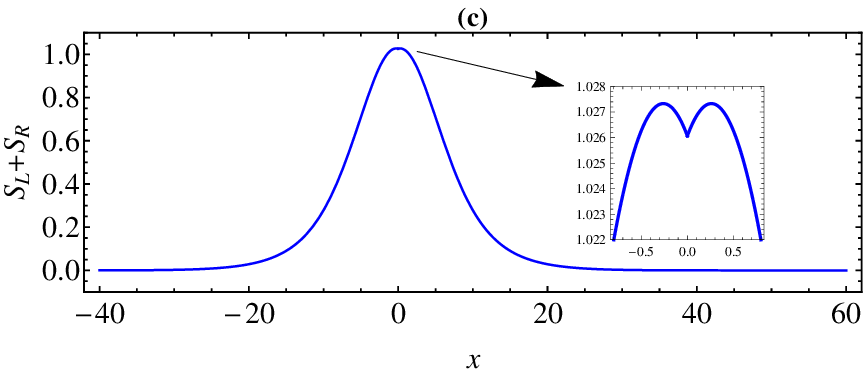}
\includegraphics[width=6cm]{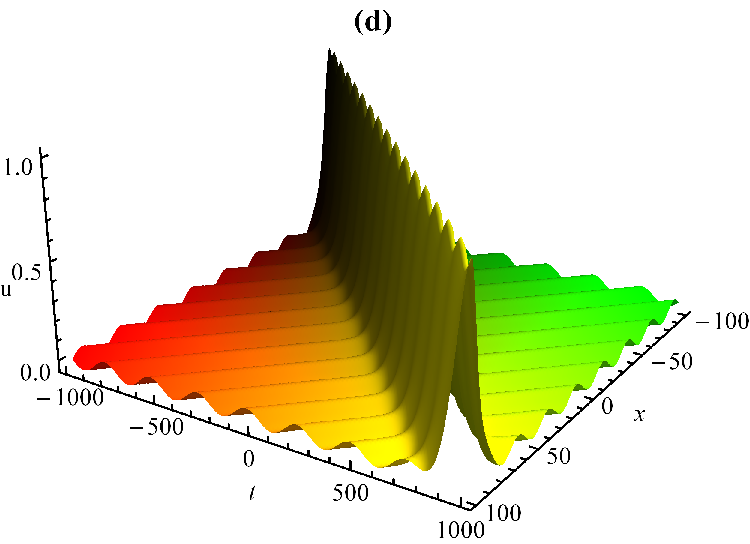}
\center{\footnotesize Fig. 4
(a) A comparison of the classical soliton solution \eqref{cls} to the nanopteron solution \eqref{solu} with $\delta=\pm1$.
(b) The related  small amplitude background wave with $\delta=1$.
(c) The related soliton structure with $\delta=1$.
(c) The three-dimensional plot of the nanopteron with $\delta=1$ for its space-time evolution.
The other parameters are $m=0.02$, $V=0.1$, $V_1=0.1$, $V_2=-0.1$, $A$ and $B$ are given by eq. \eqref{cab}.}
\end{center}
\end{figure}

\section{Summary and discussions}
In this Letter, we present a new soliton-conidal wave solution of
the KdV equation, which we also name as a nanopteron solution. Based
on this solution, some interesting features are revealed. First, it
has been observed that the soliton core preserves its shape and
velocity during the collision with the cnoidal periodic wave peaks.
Second, from the dressed structure of the solution, it is found that
the collision-induced phase shift of the cnoidal periodic wave is
always half of its wavelength, which is believed to be universal to
all the soliton-cnoidal interactions. Third, the nanopteron
structure is realized as a special limit case. It is found that for
the suitable choice of the wave parameters, the soliton core of the
soliton-cnoidal wave trends to be the classical KdV soliton and the
surrounded cnoidal periodic wave appears as small amplitude
sinusoidal variations on both sides of the main core.

The explicit solution obtained in this letter can be applied in many physical scenarios.
For instance, the nanopteron structure can be viewed as a perturbed classical soliton,
and it may provide some correction to the classical soliton in both theoretical and experimental studies.

\section{Acknowledgments}
The work was sponsored by the National Natural Science Foundations of
China (Nos. 11275123, 11175092, 11205092 and 10905038), Shanghai Knowledge Service Platform for Trustworthy Internet of Things (No. ZF1213), Scientific Research Fund of Zhejiang Provincial Education Department under Grant No. Y201017148,
and K. C. Wong Magna Fund in Ningbo University.

\end{document}